\documentclass[aps,pra,longbibliography,twocolumn,superscriptaddress]{revtex4}

\usepackage{amssymb}
\usepackage{amsmath}
\usepackage{mathdots}
\usepackage{txfonts}
\usepackage{graphicx}
\usepackage{bm}
\usepackage{color}
\usepackage{hyperref}
\usepackage{booktabs}
\usepackage{here}

\begin{document}
\title{Finite-size corrections for defect-involving vertical transitions in supercell calculations}

\author{Tomoya Gake}
\affiliation{Laboratory for Materials and Structures, Institute of Innovative Research, Tokyo Institute of Technology, Yokohama 226-8503, Japan}

\author{Yu Kumagai}
\email[]{kumagai@msl.titech.ac.jp}
\affiliation{Laboratory for Materials and Structures, Institute of Innovative Research, Tokyo Institute of Technology, Yokohama 226-8503, Japan}
\affiliation{PRESTO, Japan Science and Technology Agency, Tokyo 113-8656, Japan}

\author{Christoph Freysoldt}
\affiliation{Max-Planck-Institut f\"{u}r Eisenforschung GmbH, Max-Planck-Stra\ss e 1, 40237 D\"{u}sseldorf, Germany}

\author{Fumiyasu Oba}
\affiliation{Laboratory for Materials and Structures, Institute of Innovative Research, Tokyo Institute of Technology, Yokohama 226-8503, Japan}
\affiliation{Center for Materials Research by Information Integration, Research and Services Division of Materials Data and Integrated System, National Institute for Materials Science, Tsukuba 305-0047, Japan}

\date{\today}

\begin{abstract}
A correction method for vertical transition levels (VTLs) involving defect states calculated with a supercell technique is formulated and its effectiveness is systematically verified with ten defects in prototypical materials: cubic-BN, GaN, MgO, and 3C-SiC.
Without any corrections, the absolute errors are around 1~eV with moderate size supercells in most cases.
In contrast, when our correction method is adopted, the absolute errors are reduced and become less than 0.12 eV in all the cases.
Our correction scheme is general and will have the potential for wide application as it is adaptive for evaluating various quantities at fixed geometry, as represented by those relevant to the generalized Koopmans' theorem. 
\end{abstract}

\maketitle

Point defects determine various physical properties in solids.
One of the most important properties is the optical property.
For instance, ZnO is known to show green luminescence, which has been attributed to point defects~\cite{Rodnyi_OS2011}.
Another example is SrTiO$_{3}$, known to show blue-light emission at room temperature after Ar$^{+}$ irradiation, whose origin is presumably the emergence of oxygen vacancies~\cite{Kan_NatMat2005}.
In addition, we can investigate defects with deep states, which often degrade the device performance, from the photoabsorption and photoemission spectra.
It is generally considered that absorption (emission) of a photon by a defect promotes (demotes) an electron to the excited (ground) state, most probably without altering the atomic configuration, based on the Franck-Condon principle~\cite{Franck_TFS1926}.
This is a consequence of the fact that electrons are much lighter than nuclei.
Thus, we can represent the optical transition by a vertical arrow in the configuration coordination diagram, and its transition energy is given as an optical transition level or vertical transition level (VTL) with respect to the valence band maximum (VBM) or the conduction band minimum (CBM)~\cite{Vandewalle_JAP2004,Freysoldt_RMP2014}.

First-principles calculations have become a powerful tool to understand and predict the defect properties.
In the calculations of extended systems with defects, they are almost always evaluated by a supercell approach nowadays, where a charged defect interacts with its periodic images and background charge, which erroneously modifies the total energies of charged defect supercells~\cite{Lany_PRB2008,Freysoldt_PRL2009,Komsa_PRB2012,Kumagai_PRB2014}.
Methods to correct the energies to the dilute limit are well established, as represented by the scheme proposed by Freysoldt, Neugebauer, and Van~de~Walle (FNV)~\cite{Freysoldt_PRL2009}, and its extension to anisotropic systems and/or relaxed geometries (eFNV)~\cite{Kumagai_PRB2014}.
The correction energy of the (e)FNV scheme is written as 
\begin{equation} \label{GrindEQ__1_} 
E_{\mathrm{corr}}=E_{\mathrm{PC}}-Q\Delta V_{\mathrm{PC,}Q\mathrm{/b}}|_{\mathrm{far}}.
\end{equation} 
The first term is the point charge (PC) correction energy and the second is an alignmentlike term, where $\Delta V_{\mathrm{PC,}Q\mathrm{/b}}|_{\mathrm{far}}$ is the potential difference between defect-induced and PC potentials at a region outside of the defect in the supercell, and \textit{Q} is the defect charge state.
The PC correction energy and potential are evaluated using the static dielectric tensor because a charged defect in its ground state is screened
by both electrons and ions.
The charged defect formation energy in the dilute limit is accurately calculated using Eq.~(\ref{GrindEQ__1_}) as long as the defect charge is enclosed in a supercell~\cite{Freysoldt_PRL2009,Komsa_PRB2012,Kumagai_PRB2014}.

\begin{figure}[bt]
\includegraphics[width=1\linewidth]{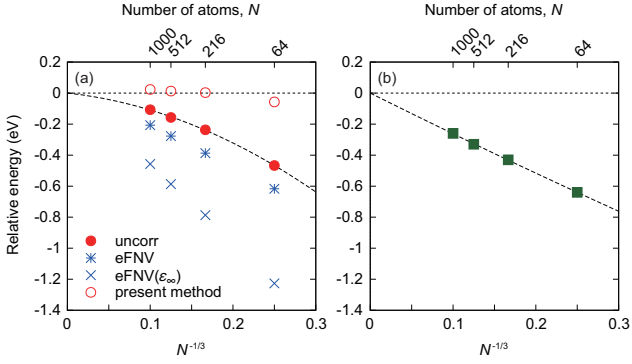}
\caption{(a) Relative VTLs from $+1$ to neutral charge transition of the oxygen vacancy (\textit{V}$_{\mathrm{O}}$) in MgO without corrections, with the eFNV corrections using the static [eFNV] or high frequency dielectric tensor [eFNV($\varepsilon_\infty$)]~\cite{Note1}, and with corrections using the present method.
(b) Relative formation energies of \textit{V}$_{\mathrm{O}}^{0}$ at the \textit{V}$_{\mathrm{O}}^{+1}$ atomic configurations in MgO without corrections.
The horizontal axis is the inverse of the cube root of the number of atoms in the supercells.
The uncorrected values are fitted with a function of $aN^{-1}+bN^{-1/3}+c$, and energy zeros are set at the respective extrapolated ones.}
\label{fig:VTL1}
\end{figure}

However, conventional corrections are not applicable to the calculation of a VTL as this involves a non-trivial combination of ionic and electronic screening, as we will explain in a moment.
In general, a vertical transition starts from some charge state \textit{Q} in the equilibrium atomic configuration \textit{S} of that charge state, \textit{S}(\textit{Q}), which includes the ionic screening response to $Q$. The VTL to final charge state $Q+\Delta Q$, involving the VBM (CBM), is calculated as 
\begin{equation} \label{GrindEQ__2_} 
\begin{split}
\mu [Q/Q+\Delta Q;S(Q)] =& \, \left\{E\left[Q+\Delta Q;S(Q)\right]-E\left[Q;S(Q)\right]\right\} \\
&+\Delta Q\cdot{\varepsilon}_{\mathrm{V(C)BM}},
\end{split}
\end{equation} 
where $\Delta Q$ is the additional charge, i.e., $+1$ or $-1$, and ${\varepsilon}_{\mathrm{V(C)BM}}$ denotes the energy level of the VBM (CBM).
Let us illustrate the issues of the VTL calculations from supercells through the example of the $+1$ to neutral charge transition of the oxygen vacancy (\textit{V}$_{\mathrm{O}}$) in MgO.
One would naively expect the VTL to be well estimated by applying the conventional FNV scheme to the initial state using the static dielectric tensor since the final state is charge neutral.
Unfortunately, this is incorrect. As shown in Fig.~\ref{fig:VTL1}(a), the eFNV-corrected VTLs vary even \emph{more} with cell size than the uncorrected ones.
As the total energy of the initial state should be well corrected by the FNV method, the cell size dependence is attributed to the energy of the final state although it is neutral.
Indeed, the formation energies of \textit{V}$_{\mathrm{O}}^{0}$ at the \textit{V}$_{\mathrm{O}}^{+1}$ atomic configurations increase with increasing supercell size as shown in Fig.~\ref{fig:VTL1}(b).
This is an artifact purely caused by a spurious electrostatic potential.
The reason is that the interaction of the added charge ($\Delta Q=-1$) with the periodic images of the initial state ($Q=+1$) are statically screened (electrons + ions), but the interactions with its own periodic images ($\Delta Q=-1$) involve electronic screening only, and the two contributions do not cancel.
Instead, one might think the (e)FNV scheme using the high-frequency (electronic contribution only) dielectric tensor works better as the energy caused by the ionic displacement is canceled out when taking the energy difference between the final and initial states.
Again, this is not correct as shown in Fig.~\ref{fig:VTL1}(a), where much worse cell size dependence remains.

Similar issues are of course present for any VTL.
Although there have been numerous theoretical reports on the calculations of VTLs~\cite{Lyons_RPL2012,npj_Lyons2017,Frodason_PRB2017,Ho_PRB2018}, surprisingly, none has referred to this issue.
Therefore, we aim to formulate a correction method for the VTLs within the total energy framework. 

To simplify the derivation, we will assume the cubic system with isotropic screening.
However, the final correction energy expression is common to the anisotropic systems.
A VTL or an additional electrostatic energy, arising from adiabatically introducing $\Delta$\textit{Q} with a spatial distribution $\Delta$\textit{q}($\boldsymbol{r}$) that excludes the screening charge, becomes for the isolated defect
\begin{equation} \label{GrindEQ__3_}
\Delta E=\int{d^3\boldsymbol{r}\Delta q\left(\boldsymbol{r}\right)V^Q\left(\boldsymbol{r}\right)}+\frac{1}{2}\int{d^3\boldsymbol{r}\Delta q\left(\boldsymbol{r}\right)V^{\Delta Q}\left(\boldsymbol{r}\right)},
\end{equation} 
where $V^Q\left(\boldsymbol{r}\right)$ is the initial defect-induced electrostatic potential of a single defect, and $V^{\Delta Q}\left(\boldsymbol{r}\right)$ an additional contribution to the electrostatic potential, which is caused by the addition of ${\Delta Q}$.
The factor 1/2 arises only for the second term as $V^{\Delta Q}\left(\boldsymbol{r}\right)$ builds up while $\Delta$\textit{Q} is introduced, whereas $V^Q\left(\boldsymbol{r}\right)$ is unchanged throughout the process.
Since \textit{Q} and $\Delta$\textit{Q} are screened with static (${\varepsilon}_0$) and electronic (${\varepsilon}_{\infty}$) dielectric constants, respectively, the asymptotic potential behaves like $\frac{Q/{\varepsilon}_0+\Delta Q/{\varepsilon}_{\infty}}{r}$ when the centers of \textit{Q} and $\Delta$\textit{Q} are assumed to be the same.
Therefore, standard finite-size supercell corrections with a single charge and a single dielectric constant, cannot capture this effect.
Also, it is not a difference of simple electrostatic corrections of initial and final state as demonstrated above.

In the dilute limit, $V^Q\left(\boldsymbol{r}\right)$ is written as $V^Q_{\mathrm{lr}}\left(\boldsymbol{r}\right)+V^Q_{\mathrm{sr}}\left(\boldsymbol{r}\right)$, where $V^Q_{\mathrm{lr}}\left(\boldsymbol{r}\right)$ is the long-range potential caused by point charge \textit{Q} and $V^Q_{\mathrm{sr}}\left(\boldsymbol{r}\right)$ the remaining part representing the short-range potential.
Similarly, $V^{\Delta Q}\left(\boldsymbol{r}\right)$ is written as $V^{\Delta Q}_{\mathrm{lr}}\left(\boldsymbol{r}\right)+V^{\Delta Q}_{\mathrm{sr}}\left(\boldsymbol{r}\right)$.
Following the discussion in Ref.~\cite{Freysoldt_PRL2009}, the short-range potential can be obtained directly from the density functional theory (DFT) electrostatic potentials $\tilde{V}^{\mathrm{DFT}}$ as
\begin{eqnarray} \label{GrindEQ__4_}
V^Q_{\mathrm{sr}}\left(\boldsymbol{r}\right) &=& {\tilde{V}}\mathrm{^{DFT}_{initial}}\left(\boldsymbol{r}\right)-{\tilde{V}}\mathrm{^{DFT}_{bulk}}\left(\boldsymbol{r}\right)-{\tilde{V}}^Q_{\mathrm{lr}}\left(\boldsymbol{r}\right)-C^Q, \\
V^{\Delta Q}_{\mathrm{sr}}\left(\boldsymbol{r}\right) &=& {\tilde{V}}\mathrm{^{DFT}_{final}}\left(\boldsymbol{r}\right)-{\tilde{V}}\mathrm{^{DFT}_{initial}}\left(\boldsymbol{r}\right)-{\tilde{V}}^{\Delta Q}_{\mathrm{lr}}\left(\boldsymbol{r}\right)-C^{\Delta Q}.
\end{eqnarray} 
The symbol $\tilde{V}$ is used for periodic potentials defined within the DFT supercell.
The initial state is the defect-containing supercell in charge state \textit{Q}, while the final state is the one in charge state $Q+\Delta Q$.
$C^Q$ and $C^{\Delta Q}$ are alignment constants chosen such that the short-range potentials decay to zero far from the defect~\cite{Freysoldt_PRL2009,Freysoldt_PSSB2011}.
The correction potentials for charge \textit{Q} and $\Delta$\textit{Q} are therefore written as
\begin{eqnarray} \label{GrindEQ__5_}
V^Q_\mathrm{corr}\left(\boldsymbol{r}\right) &=& \left[V^Q_{\mathrm{lr}}\left(\boldsymbol{r}\right)+V^Q_{\mathrm{sr}}\left(\boldsymbol{r}\right)\right]-\left[{\tilde{V}}\mathrm{^{DFT}_{initial}}\left(\boldsymbol{r}\right)-{\tilde{V}}\mathrm{^{DFT}_{bulk}}\left(\boldsymbol{r}\right)\right] \nonumber \\
&=& V^{Q,\mathrm{corr}}_{\mathrm{lr}}\left(\boldsymbol{r}\right)-C^Q, \\
V^{\Delta Q}_\mathrm{corr}\left(\boldsymbol{r}\right) &=& \left[V^{\Delta Q}_{\mathrm{lr}}\left(\boldsymbol{r}\right)+V^{\Delta Q}_{\mathrm{sr}}\left(\boldsymbol{r}\right)\right]-\left[{\tilde{V}}\mathrm{^{DFT}_{final}}\left(\boldsymbol{r}\right)-{\tilde{V}}\mathrm{^{DFT}_{initial}}\left(\boldsymbol{r}\right)\right] \nonumber \\
&=& V^{\Delta Q,\mathrm{corr}}_{\mathrm{lr}}\left(\boldsymbol{r}\right)-C^{\Delta Q},
\end{eqnarray} 
where $V^{Q,\mathrm{corr}}_{\mathrm{lr}}\left(\boldsymbol{r}\right)=V^Q_{\mathrm{lr}}\left(\boldsymbol{r}\right)-{\tilde{V}}^Q_{\mathrm{lr}}\left(\boldsymbol{r}\right)$ and $V^{\Delta Q,\mathrm{corr}}_{\mathrm{lr}}\left(\boldsymbol{r}\right)=V^{\Delta Q}_{\mathrm{lr}}\left(\boldsymbol{r}\right)-{\tilde{V}}^{\Delta Q}_{\mathrm{lr}}\left(\boldsymbol{r}\right)$ mean the sign-reversed Madelung potentials.
The correction energy is then obtained by replacing the electrostatic potentials to the correction potentials
\begin{equation} \label{GrindEQ__6_}
\Delta E_\mathrm{corr}=\ \int{d^3\boldsymbol{r}\Delta q\left(\boldsymbol{r}\right)V^Q_\mathrm{corr}\left(\boldsymbol{r}\right)}+\frac{1}{2}\int{d^3\boldsymbol{r}\Delta q\left(\boldsymbol{r}\right)V^{\Delta Q}_\mathrm{corr}\left(\boldsymbol{r}\right)}.
\end{equation} 
Here, we split the additional charge $\Delta q\left(\boldsymbol{r}\right)$ to ${\rho}^{\Delta Q}_\mathrm{PC}$ and ${\Delta\rho}^{\Delta Q}$, where ${\rho}^{\Delta Q}_\mathrm{PC}$ is the point charge $\Delta Q$ at the center of the defect and ${\Delta\rho}^{\Delta Q}$ is the finite-size distribution.
Based on the conventional FNV scheme, the second term in Eq.~(\ref{GrindEQ__6_}) is written as $\frac{\alpha{\left(\Delta Q\right)}^2}{2{\varepsilon}_{\infty}L}-{\Delta QC}^{\Delta Q}$~\cite{Freysoldt_PRL2009,Komsa_PRB2012}, where \textit{L} and $\alpha$ are the size of the supercell and the Madelung constant, respectively.
In the same manner, the first term is rewritten as $\frac{\alpha Q\Delta Q}{{\varepsilon}_0L}-{\Delta QC}^Q-\int{d^3\boldsymbol{r}}{\Delta\rho}^{\Delta Q}V^{Q}_\mathrm{corr}\left(\boldsymbol{r}\right)$.
Here, the third term indicates the interaction between the finite size of the additional charge and the correction potential for the initial charge.
When we assume Coulomb interactions are screened by a dielectric constant independently of distance, $V^{Q}_\mathrm{corr}\left(\boldsymbol{r}\right)$ can be expanded as,
\begin{equation} \label{GrindEQ__7_}
V^{Q}_\mathrm{corr}\left(\boldsymbol{r}\right)=\frac{\alpha Q}{{\varepsilon}_0L}-\frac{2\pi \int{d^3\boldsymbol{r}}q\left(\boldsymbol{r}\right)r^2}{3{\varepsilon}_0L^3}-\frac{2\pi Qr^2}{3{\varepsilon}_0L^3}.
\end{equation} 
\begin{figure*}[t]
\includegraphics[width=1\linewidth]{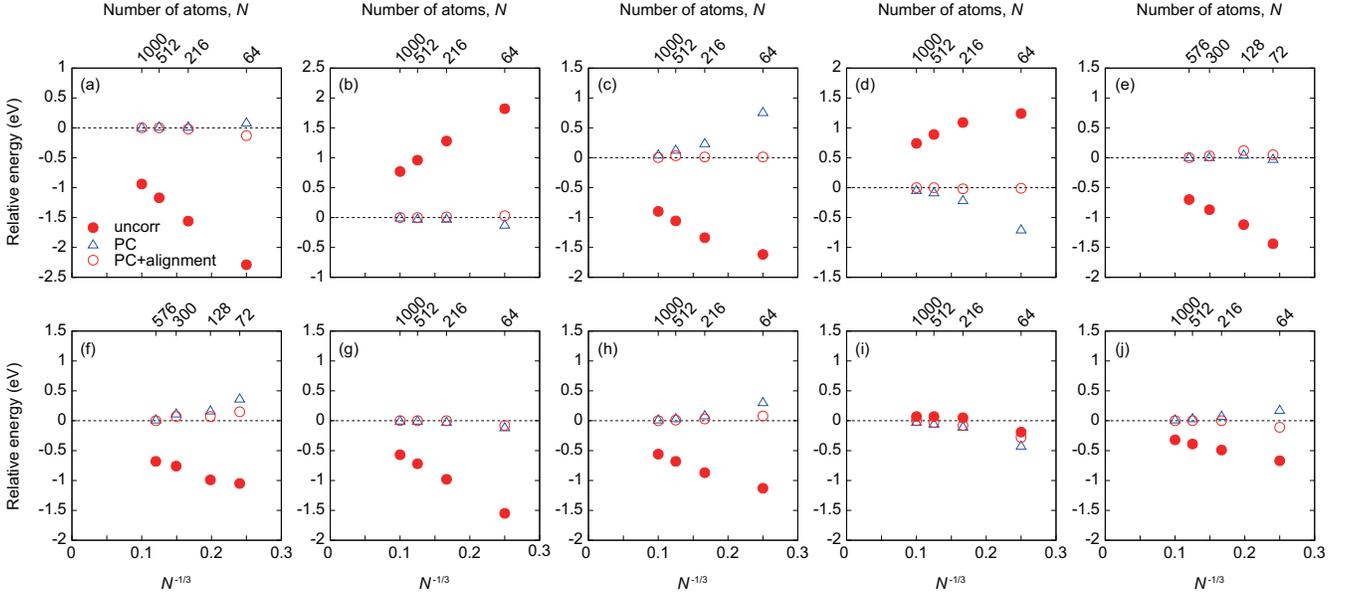}
\caption{Relative VTLs without corrections, with point-charge (PC) corrections, and with PC plus alignment corrections of (a) \textit{V}$_{\mathrm{B}}$($-2$/$-3$), (b) \textit{V}$_{\mathrm{B}}$($-3$/$-2$), (c) \textit{V}$_{\mathrm{N}}$($+2$/$+3$), and (d) \textit{V}$_{\mathrm{N}}$($+3$/$+2$) in c-BN; (e) \textit{V}$_{\mathrm{Ga}}$($-2$/$-3$) and (f) \textit{V}$_{\mathrm{N}}$($+2$/$+3$) in GaN; (g) \textit{V}$_{\mathrm{Mg}}$($-1$/$-2$), (h) \textit{V}$_{\mathrm{O}}$($+1$/$+2$), and (i) \textit{V}$_{\mathrm{O}}$($+2$/$+1$) in MgO; and (j) \textit{V}${}_{\mathrm{C}}$($+1$/$+2$) in 3C-SiC as a function of the inverse of the cube root of the number of atoms in the supercells.
The energy zeros are set at the values with the PC plus alignment corrections at the largest supercells.}
\label{fig:VTL2}
\end{figure*}
Using
\begin{equation} \label{GrindEQ__8_}
C^{\Delta Q}=\frac{2\pi \int{d^3\boldsymbol{r}}{\Delta\rho}^{\Delta Q}\left(\boldsymbol{r}\right)r^2}{3{{\varepsilon}_{\infty}L}^3}
\end{equation} 
as derived by Komsa \textit{et al}.~\cite{Komsa_PRB2012}, 
\begin{equation} \label{GrindEQ__9_}
\int{d^3\boldsymbol{r}}{\Delta\rho}^{\Delta Q}V^{Q}_\mathrm{corr}\left(\boldsymbol{r}\right)={\frac{{\varepsilon}_{\infty}}{{\varepsilon}_0}QC}^{\Delta Q}.
\end{equation} 
By summing all the contributions, the correction energy can be written as
\begin{equation} \label{GrindEQ__10_}
\begin{split}
\Delta E_\mathrm{corr} =&\, \frac{\alpha Q\Delta Q}{{\varepsilon}_0L}+\frac{\alpha{\left(\Delta Q\right)}^2}{2{\varepsilon}_{\infty}L} \\
&-\left(\Delta QC^{\Delta Q}+\Delta QC^Q+\frac{{\varepsilon}_{\infty}}{{\varepsilon}_0}QC^{\Delta Q}\right).
\end{split}
\end{equation} 
For anisotropic systems, one needs to use the averaged dielectric constant, i.e., $\langle\varepsilon\rangle=\frac{1}{3}\left(\varepsilon_{\mathrm{11}}+\varepsilon_{\mathrm{22}}+\varepsilon_{\mathrm{33}}\right)$, where $\varepsilon_{ij}$ is a component of dielectric tensor $\bar{\varepsilon}$.
Then, the correction energy can be rewritten as
\begin{equation} \label{GrindEQ__11_}
\begin{split}
\Delta E_\mathrm{corr} =&\, \frac{2\Delta Q}{Q} E^Q_\mathrm{PC}\left(\bar{\varepsilon}_0\right)+E^{\Delta Q}_{\mathrm{PC}}\left(\bar{\varepsilon}_\infty\right) \\
&-\left(\Delta QC^{\Delta Q}+\Delta QC^Q+\frac{\left\langle{\varepsilon}_{\infty}\right\rangle}{\left\langle{\varepsilon}_0\right\rangle}QC^{\Delta Q}\right),
\end{split}
\end{equation} 
where $E^{Q}_\mathrm{PC}\left(\bar{\varepsilon}\right)$ is the point-charge corrections for charge \textit{Q} screened by $\bar{\varepsilon}$.
The first two contributions mean first-order point-charge correction, while the third term (in parentheses) corresponds to the alignment term. 

To test our correction method, we investigate the supercell size dependence of the VTLs involving defects in prototypical nonmetallic materials, namely cubic-BN (c-BN), GaN, MgO, and 3C-SiC, with respect to the CBM and VBM [Eq.~(\ref{GrindEQ__2_})] for donor- and acceptor-type defects, respectively.
All the VTLs in this study were calculated with the Heyd-Scuseria-Ernzerhof hybrid functional~\cite{Heyd_JCP2003}, and the Fock-exchange parameters were tuned to reproduce each experimental band gap~\cite{Varley_PRB2012,Gordon_PRB2015,npj_Lyons2017,Weston_PRB2017}.
We also confirmed that the formation energies of all the defects considered in this study were accurately corrected with eFNV (see Fig. S1 in the Supplemental Material~\cite{Note2}), which indicated these defects were mostly enclosed by relatively smaller, e.g., 64-atom supercells. 

Figures~\ref{fig:VTL2}(a) and \ref{fig:VTL2}(b) show both the corrected and uncorrected VTLs of \textit{V}$_{\mathrm{B}}$ in c-BN as a function of the supercell size.
Without any corrections, one can see the VTLs strongly depend on the supercell size, and the absolute error is larger than 0.7 eV even with the largest 1000-atom supercell in our test set.
Conversely, the PC-level correction based on Eq. (\ref{GrindEQ__11_}) drastically reduces the cell size dependences, and the absolute errors are only 0.1 eV at the smallest 64-atom supercell.
Such a small cell size dependence indicates the \textit{V}$_{\mathrm{B}}^{-2}$ and \textit{V}$_{\mathrm{B}}^{-3}$ defect charges are very much localized.
Indeed, their defect formation energies are satisfactorily corrected with the PC corrections as well.

Figure~\ref{fig:VTL2} also shows the rest of the VTLs involving vacancies in the four materials as a function of the supercell size.
Again, our PC corrections work well with reasonable accuracy on average.
Regarding cation vacancies [Figs.~\ref{fig:VTL2}(e) and \ref{fig:VTL2}(g)], the corrected values are almost constant within 0.1 eV even at the smallest supercells, as in the case of \textit{V}$_{\mathrm{B}}$ in c-BN [Figs.~\ref{fig:VTL2}(a) and \ref{fig:VTL2}(b)].
In contrast, the VTLs of anion vacancies are not so well corrected especially at the smallest supercells [Figs.~\ref{fig:VTL2}(c), \ref{fig:VTL2}(d), \ref{fig:VTL2}(f), and \ref{fig:VTL2}(h)--\ref{fig:VTL2}(j)]: the worst case is \textit{V}$_{\mathrm{N}}$($+2$/$+3$) in c-BN with a deviation of 0.8 eV at the 64-atom supercell compared with the corrected VTL at the 1000-atom supercell [Fig.~\ref{fig:VTL2}(c)].
This means the PC corrections are not appropriate for the anion vacancies, which is consistent with the fact that their total energies are not corrected by the PC corrections (see Fig. S1 in the Supplemental Material~\cite{Note2}).

As shown in Fig.~\ref{fig:VTL2}, corrections with the alignment term successfully estimate the VTLs in the dilute limit even with relatively smaller supercells and work better than the PC corrections in most cases.
Notable examples are \textit{V}$_{\mathrm{N}}$($+2$/$+3$) and \textit{V}$_{\mathrm{N}}$($+3$/$+2$) in c-BN [Figs.~\ref{fig:VTL2}(c) and \ref{fig:VTL2}(d)]; the PC corrections lead to errors of 0.8 and 0.7 eV at the smallest supercells, respectively, while the errors are reduced to 0.0 eV when the alignment terms are incorporated.
Exceptions are \textit{V}$_{\mathrm{B}}$($-2$/$-3$) in c-BN [Fig.~\ref{fig:VTL2}(a)] and \textit{V}$_{\mathrm{Ga}}$($-2$/$-3$) in GaN [Fig.~\ref{fig:VTL2}(e)], where the alignment terms slightly increase the errors.

\begin{figure}[bt]
\includegraphics[width=0.9\linewidth]{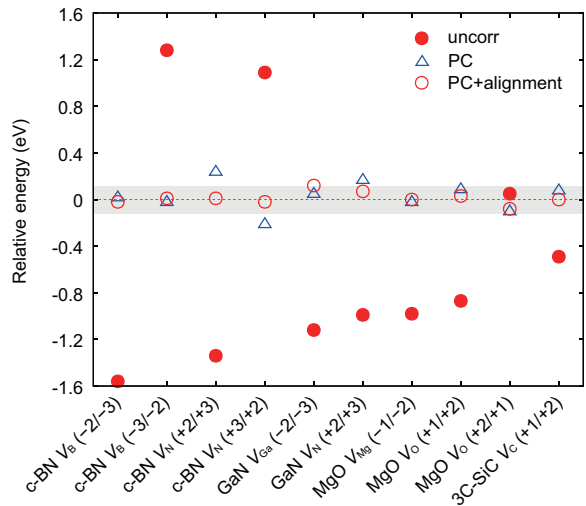}
\caption{Relative VTLs without corrections, with PC corrections, and with PC plus alignment corrections estimated with the second-smallest supercells (128 atoms for GaN and 216 atoms for c-BN, MgO, and 3C-SiC).
The energy zeros are the same as that in Fig.~\ref{fig:VTL2}. The shaded area within $\pm$0.12 eV highlights the errors of the PC plus alignment corrections.} 
\label{fig:VTL3}
\end{figure}

To see the overall performance of our correction methods, we align the VTLs calculated with the second-smallest supercells (128 atoms for GaN and 216 atoms for the others) to the same scale in Fig.~\ref{fig:VTL3}; such moderate size supercells are regularly adopted for practical defect calculations nowadays.
Although the VTLs are reasonably well corrected even at the PC correction level, the addition of the alignment term reduces the errors to be within 0.12 eV in all the cases.   
Another important merit to consider for the alignmentlike term is that it removes the projector augmented wave (PAW) or pseudopotential core radius dependences from the VTLs.
Bruneval and co-workers have reported that the charged defect formation energies depend on the PAW or pseudopotential radii, because the standard of the electrostatic potential depends on the radii, and this modifies the total energy of a charged system~\cite{Bruneval_PRB2014}.
The VTLs depend on these radii as well because the charge state is different by $+1$ or $-1$ between the final and initial states, but such an energy dependence is removed when an alignmentlike term, more specifically the second term in the parentheses in Eq.~(\ref{GrindEQ__11_}), is considered.

Finally, we emphasize our correction method is adaptive for evaluating any quantities involving different charge states at fixed geometry.
One of the most important examples is the quantity relevant to the generalized Koopmans' theorem (gKT)~\cite{Perdew_PRL1982}.
On the basis of the gKT, the deviation from linearity is evaluated at a fixed geometry as $\Delta_{\mathrm{XC}}$:
\begin{equation} \label{GrindEQ__12_} 
\Delta_{\mathrm{XC}}=\{E\left[Q+\Delta Q;S(Q)\right]-E\left[Q;S(Q)\right]\}+\Delta Q\cdot \varepsilon\left[Q;S(Q)\right], 
\end{equation} 
where $\varepsilon[Q;S(Q)]$ is a single-particle level of a localized orbital such as a defect-induced state.
$\Delta_{\mathrm{XC}}$ has recently been actively used to evaluate the accuracy of the exchange-correlation functional for studying point defects and small polarons~\cite{Miceli_PRB2018,Elmaslmane_JCTC2018,Gake_PRM2019}.
The finite size correction to the eigenvalue for the second term in Eq. (\ref{GrindEQ__12_}) has been well established by many authors~\cite{Jain_PRL2011,Komsa_PRB2012,Chen_PRB2013}.
Conversely, the correction to the first bracketed term has not been discussed so far.
This is indeed the same as Eq.~(\ref{GrindEQ__2_}), and therefore our newly developed correction is applicable for it.
We note here that the eigenvalues are not corrected as accurately as the total energies or the VTLs as shown in Fig. S2 in the Supplemental Material~\cite{Note2}.
Therefore, we recommend the use of larger supercells when one needs to accurately determine $\Delta_{\mathrm{XC}}$.

In summary, we derived correction schemes for calculating VTLs using defect supercells, and tested the effectiveness with ten vacancies in prototypical materials: cubic-BN, GaN, MgO, and 3C-SiC.
At the smallest supercells, the PC corrections satisfactorily evaluate the VTLs in the dilute limit involving cation vacancies but do not work well for those involving anion vacancies.
We then verified that potential alignment terms reduce the absolute errors to less than 0.12 eV at the second-smallest supercells in our test set.
Since our method is routinely and automatically applied, systematic calculations of VTLs are feasible.
Furthermore, it is adaptive for evaluating any quantities involving different charge states at fixed geometry, as represented by the gKT-relevant quantity.

\textit{Computational methods}.
First-principles calculations were performed using the PAW method~\cite{Blochl_PRB1994}, as implemented in VASP~\cite{Kresse_PRB1996,Kresse_PRB1999}.
We used PAW data sets with radial cutoffs of 0.91, 1.40, 1.52, 1.31, 0.74, 0.90, and 0.86 $\text{\AA}$ for B, Ga, Mg, Si, N, O, and C, respectively.
In the supercell calculations, a 2~$\times$~2~$\times$~2 \textit{k}-point mesh was applied to the 216-atom or smaller supercells, while $\Gamma$-only sampling was applied to the others, and spin polarization was considered in all the defect calculations.
Note that all the VTLs considered did not involve partial occupation at the VBM, CBM or defect localized states.

\textit{Acknowledgments}.
Fruitful discussions with Naoki Tsunoda are deeply appreciated.
This work was supported by the Grants-in-Aid for Scientific Research B (Grant No. 19H02416), PRESTO (Grant No. JPMJPR16N4), CREST (Grant No. JPMJCR17J2), and the Support Program for Starting Up Innovation Hub MI$^2$I from JST, Japan.
The computing resources of ACCMS at Kyoto University and Research Institute for Information Technology at Kyushu University were used for part of this work.

T.G. and Y.K. contributed equally to this work.

\end{document}


\title{\Huge Supplemental Material}

\author[1]{Tomoya Gake}
\author[1,2]{Yu Kumagai}
\author[3]{Christoph Freysoldt}
\author[1,4]{Fumiyasu Oba}

\affil[1]{\it Laboratory for Materials and Structures,  Institute of Innovative Research, Tokyo Institute of Technology, Yokohama 226-8503, Japan}
\affil[2]{\it PRESTO, Japan Science and Technology Agency, Tokyo 113-8656, Japan}
\affil[3]{\it Max-Planck-Institut f\"{u}r Eisenforschung GmbH, Max-Planck-Stra\ss e 1, 40237 D\"{u}sseldorf, Germany}
\affil[4]{\it Center for Materials Research by Information Integration, Research and Services Division of Materials Data and Integrated System, National Institute for Materials Science, Tsukuba 305-0047, Japan.}

\date{\today}


\definecolor{blue}{rgb}{0.0,0.0,1}
\hypersetup{colorlinks,breaklinks,
            linkcolor=blue,urlcolor=blue,
            anchorcolor=blue,citecolor=blue}

\maketitle
\clearpage

\begin{figure}
  \centering
  \includegraphics[width=1\linewidth]{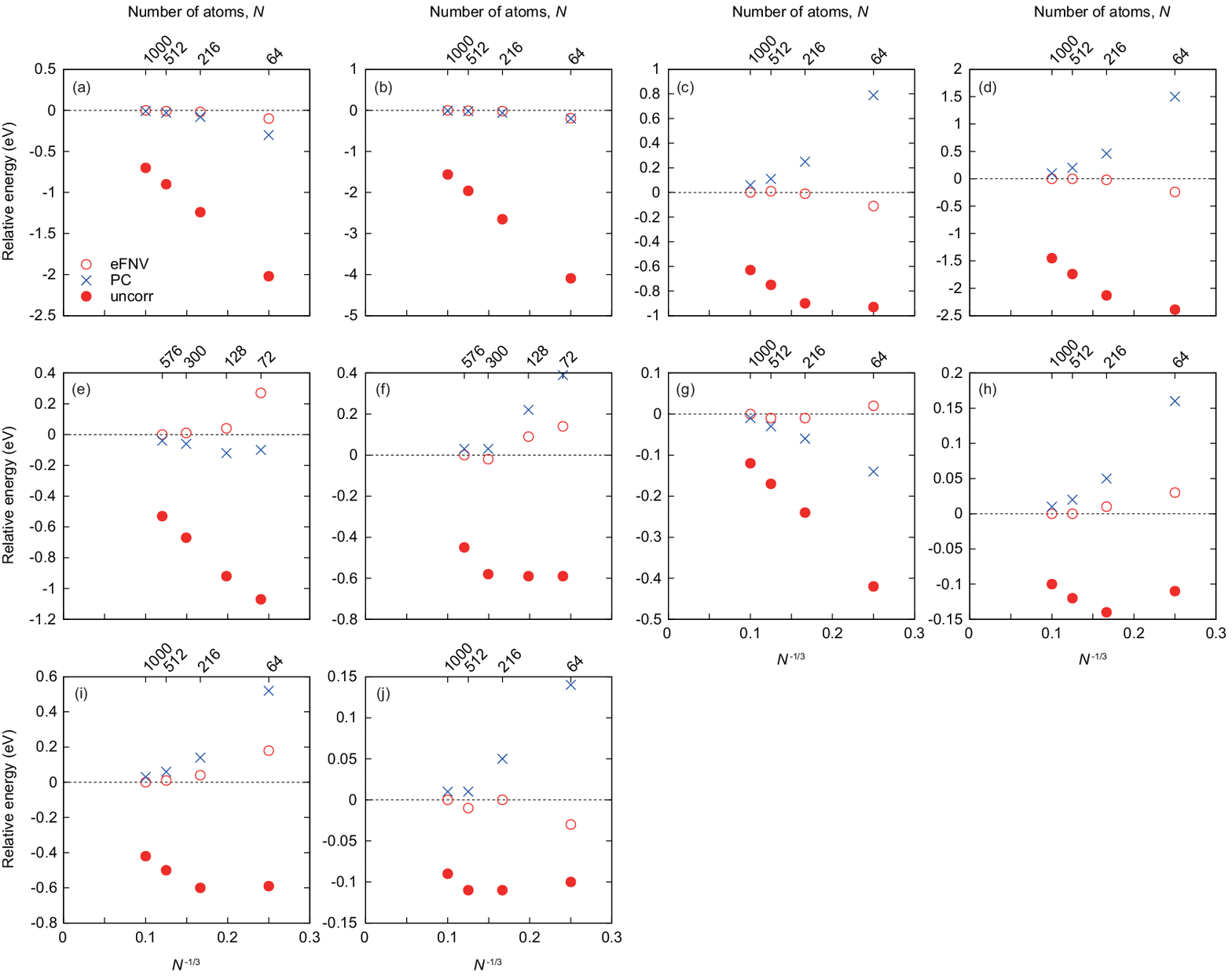}
\caption{Relative formation energies of (a) \textit{V}$_{\mathrm{B}}^{-2}$, (b) \textit{V}$_{\mathrm{B}}^{-3}$, (c) \textit{V}$_{\mathrm{N}}^{+2}$, and (d) \textit{V}$_{\mathrm{N}}^{+3}$ in c-BN; (e) \textit{V}$_{\mathrm{Ga}}^{-2}$ and (f) \textit{V}$_{\mathrm{N}}^{+2}$ in GaN; (g) \textit{V}$_{\mathrm{Mg}}^{-1}$, (h) \textit{V}$_{\mathrm{O}}^{+1}$, and (i) \textit{V}$_{\mathrm{O}}^{+2}$ in MgO; and (j) \textit{V}$_{\mathrm{C}}^{+1}$ in 3C-SiC as a function of the supercell size.
The energy zeros are set at the corrected values with the eFNV scheme for the largest supercells.} 
  \label{}
\end{figure}
\clearpage

\begin{figure}
  \centering
  \includegraphics[width=1\linewidth]{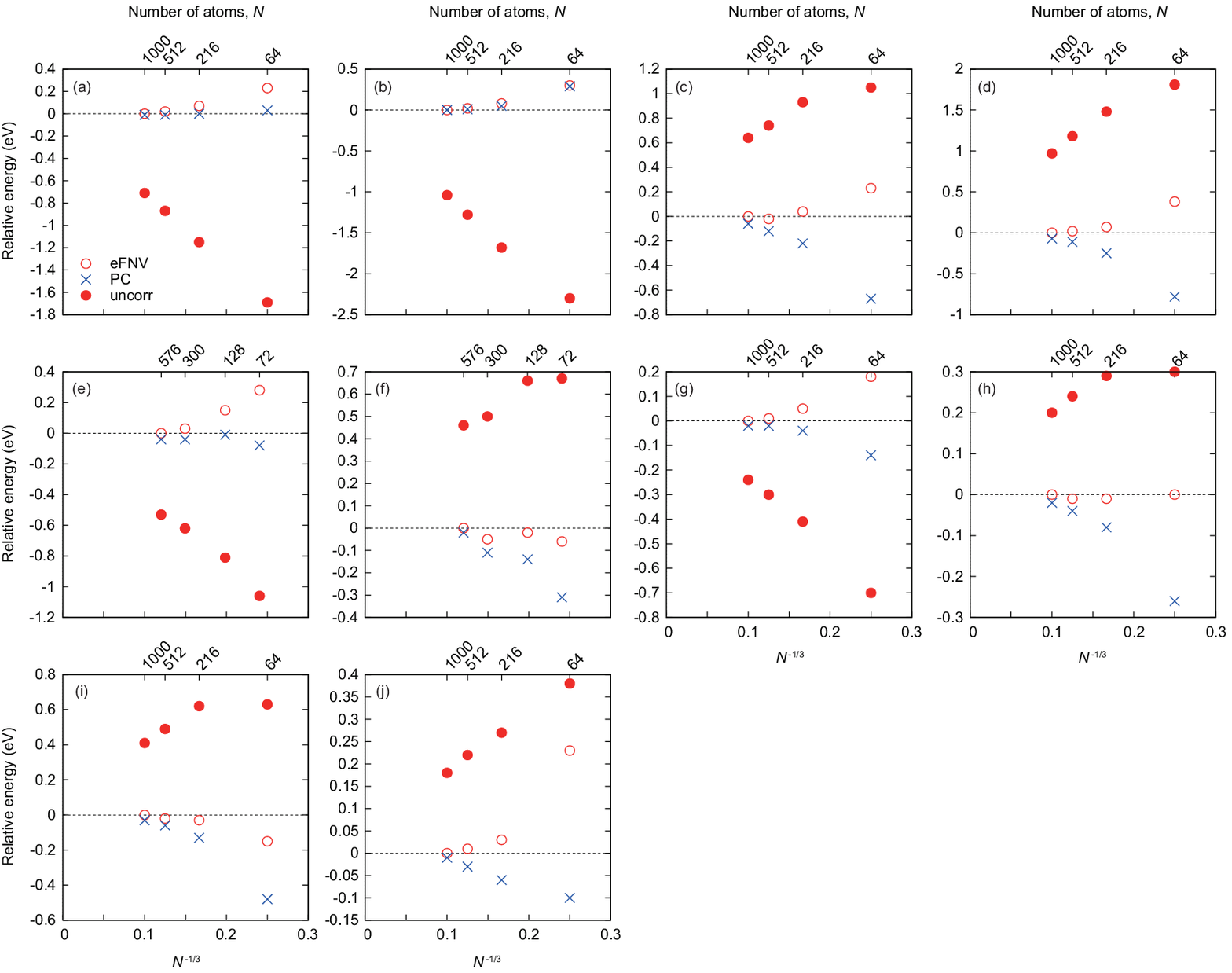} 
\caption{Relative single-particle energy levels corrected by adding $-\mathrm{2}E_{\mathrm{corr}}\mathrm{/}Q$, where $E_{\mathrm{corr}}$ is the total energy correction and \textit{Q} the charge state, of (a) \textit{V}$_{\mathrm{B}}^{-2}$, (b) \textit{V}$_{\mathrm{B}}^{-3}$, (c) \textit{V}$_{\mathrm{N}}^{+2}$, and (d) \textit{V}$_{\mathrm{N}}^{+3}$ in c-BN; (e) \textit{V}$_{\mathrm{Ga}}^{-2}$ and (f) \textit{V}$_{\mathrm{N}}^{+2}$ in GaN; (g) \textit{V}$_{\mathrm{Mg}}^{-1}$, (h) \textit{V}$_{\mathrm{O}}^{+1}$, and (i) \textit{V}$_{\mathrm{O}}^{+2}$ in MgO; and (j) \textit{V}$_{\mathrm{C}}^{+1}$ in 3C-SiC as a function of the supercell size.
The energy zeros are set at the corrected values with the eFNV correction energies for the largest supercells.}
  \label{}
\end{figure}